\documentclass[aps,prb,twocolumn,10pt]{revtex4-2}
\usepackage[utf8]{inputenc}
\usepackage{libertine}
\usepackage[T1]{fontenc}
\usepackage[libertine,upint]{newtxmath} 
\usepackage{graphicx,microtype,mathtools,siunitx,bm,expl3}
\usepackage[cal=boondoxo,frak=boondox]{mathalfa}
\usepackage[colorlinks=true,allcolors=blue]{hyperref}

\DeclareSIUnit\angstrom{\text{Å}}
\DeclarePairedDelimiter\av{\langle}{\rangle}
\DeclareMathOperator{\sign}{sign}
\DeclareMathOperator{\re}{Re}
\DeclareMathOperator{\im}{Im}

\NewDocumentCommand{\normord}{m}{%
    :\mathrel{\mspace{1mu} #1 \mspace{1mu}}:%
}

\NewDocumentCommand{\grad}{e{_^}}{%
    \mathop{}\!
    \nabla
    \IfValueT{#1}{_{\!#1}}
    \IfValueT{#2}{^{#2}}
} 

\makeatletter
\g@addto@macro\@floatboxreset\centering
\makeatother

\begin{document}

\title{Electron correlations due to pair spin-orbit interaction in 2D electron systems}

\author{Yasha Gindikin and Vladimir A.\ Sablikov}
\affiliation{Kotelnikov Institute of Radio Engineering and Electronics, Russian Academy of Sciences, Fryazino branch, Fryazino, 141190, Russia}

\begin{abstract}
We show that the pair spin-orbit interaction (PSOI) created by the Coulomb fields of interacting electrons leads to the appearance of an unconventional correlated electronic state in two-dimensional materials with the large Rashba effect. The effect manifests itself at sufficiently strong PSOI as a sharp peak in the structure factor, indicating a tendency to form a striped structure with a spatial scale determined by the competition between the Coulomb repulsion and the PSOI--induced attraction of electrons. Above the critical value of PSOI, the system becomes unstable with respect to the charge density fluctuations on this scale. 
\end{abstract}
 
\maketitle 
\section{Introduction}
The interplay of the strong spin-orbit interaction (SOI) in Rashba materials~\cite{manchon2015new} and Coulomb electron-electron (e-e) repulsion leads to a bunch of qualitatively new physical effects, including the emergence of new correlated states~\cite{PhysRevB.85.035116,PhysRevB.89.155103,PhysRevB.88.075115,PhysRevLett.115.026401}, unusual collective modes~\cite{PhysRevB.91.035106,D_Amico_2019,PhysRevB.102.195208}, and even bound electron pairs (BEPs)~\cite{PhysRevB.104.125103}.

The Rashba SOI is produced by the electric fields external to the crystal lattice. In quantum structures the common sources of this field include the confining potential, the charged impurities and structure defects. It is well known that the Coulomb fields of interacting electrons also produce the SOI which manifests directly in the e-e interaction Hamiltonian~\cite{bethe2012quantum}. As a result the interaction Hamiltonian gains the contribution that depends on the electron spins and momenta. Effects of this so-called pair spin-orbit interaction (PSOI) were until very recently considered only as a small perturbation in such problems as spin dynamics, spin-spin interaction, spin current generation, etc~\cite{BOGUSLAWSKI1980389,PhysRevB.72.161304,PhysRevB.79.195305,PhysRevB.84.033305}. However, similarly to the Rashba SOI, the PSOI is strongly enhanced in Rashba materials and therefore can produce strong changes in electronic states~\cite{2019arXiv190506340G}. 

The strong PSOI can generate a plethora of non-trivial effects due to the effective attraction that this interaction creates. The attraction mechanism is quite clear~\cite{2019arXiv190506340G}. The PSOI created by the Coulomb electric field of a given electron decreases the energy of another electron possessing a particular spin orientation relative to its momentum. This effect increases with decreasing the distance between electrons, which exactly implies the attraction. The attraction can lead to the emergence of BEPs with highly unusual configuration of the charge and spin density~\cite{2018arXiv180410826G,PhysRevB.98.115137,10.1016/j.physe.2018.12.028,2019arXiv190409510G,bhzbeps}. 

Of greatest interest is, of course, the collective behavior of a many-electron system with a strong PSOI, but this problem has been studied extremely poorly to date. In Ref.~\cite{PhysRevB.95.045138} we considered a specific situation of a gated one-dimensional quantum wire with the PSOI produced by means of the image charges on the gate. In this case the PSOI leads to the appearance of a correlated state with unusual collective excitations. One of the two collective modes strongly softens in the long-wavelength part of the spectrum with increasing the PSOI strength, and becomes unstable when the PSOI exceeds a critical value.

In this paper, we turn to a more general statement of the problem by considering a two-dimensional (2D) electron system with a richer configuration of Coulomb fields generating PSOI\@. The electron dynamics is described using the $k \cdot p$ method~\cite{voon2009kp}, assuming that the Coulomb electric fields are sufficiently smooth.

To begin with, we note that in the presence of the PSOI the effective strength of the e-e interaction is determined by two parameters. This is in stark contrast to the conventional case of an electron gas with Coulomb interaction only, where the interaction strength is characterized by the parameter $r_s$, which is the ratio of the inter-electron distance to the Bohr radius $a_B$. The e-e interaction Hamiltonian contains the PSOI component $H_{\mathrm{PSOI}}$ in addition to the usual Coulomb term $H_{\mathrm{Coul}}$, so that the system Hamiltonian is
\begin{equation}
\label{ham}
    \begin{split}
    H ={} & H_{\mathrm{kin}} + H_{\mathrm{Coul}} + H_{\mathrm{PSOI}}\\
        ={} &  \sum_{i} \frac{\hat{\bm{p}}_i^2}{2m} + \frac12 \sum_{i \ne j} \mathcal{U} (\bm{r}_{i} - \bm{r}_{j}) \\
        &{}+ \frac{\alpha}{\hbar} \sum_{i \ne j} \left( \hat{\bm{p}}_{i} \times \bm{\mathcal{E}} (\bm{r}_{i} - \bm{r}_{j})  \right) \cdot \bm{\sigma}_{i} \,.
    \end{split}
\end{equation}
Here $\mathcal{U}(\bm{r}) = e^2/ \epsilon r$ is the Coulomb interaction potential, $\bm{\mathcal{E}}(\bm{r}) = \frac{1}{e} \grad_{\bm{r}} \mathcal{U}(\bm{r})$ is the pair Coulomb field that produces PSOI, $\hat{\bm{p}}_i$ is the momentum operator of the $i$-th electron, $m$ is the effective mass, $\bm{\sigma} \equiv (\sigma_x,\sigma_y,\sigma_z)$ is the Pauli vector, and $\alpha$ stands for the Rashba constant, which is estimated for the currently known materials with strong SOI from $\SI{e2}{e \angstrom^2}$ in $\mathrm{InAs}$ to $\SI{e3}{e \angstrom^2}$ in such materials as $\mathrm{Bi}_2 \mathrm{Se}_3$~\cite{PhysRevLett.107.096802}, the $\mathrm{BiSb}$ monolayers~\cite{PhysRevB.95.165444}, and Van der Waals materials with adsorbed heavy elements~\cite{otrokov2018evidence,PhysRevB.99.085411}. In what follows it is convenient to introduce the dimensionless Rashba constant $\tilde{\alpha} = \alpha/e a_B^2$. 

The e-e interaction strength is characterized by the ratio of the interaction energy to the Fermi energy. The parameter $r_s$ relates only to the Coulomb term. The contribution of the PSOI term is described by another parameter $\tilde{\alpha}/r_s$. It is remarkable that both parameters depend differently on the parameters of the electronic system. In particular, while the parameter $r_s$ decreases with increasing the electron density, the parameter $\tilde{\alpha}/r_s$, on the contrary, increases. Therefore the PSOI correlations can dominate when the density is high enough. While the effect of Coulomb correlations is largely understood~\cite{giuliani2005quantum}, the role of the PSOI-induced correlations and the conditions under which they lead to a radical rearrangement of the electronic system remain to be elucidated. 

This paper aims to find out whether the PSOI creates characteristic correlations, under what conditions they become significant, and how this manifests itself in the spectrum of collective excitations. To this end, we study the collective excitations and charge correlations in a 2D electron system with the in-plane reflection symmetry, where the PSOI is produced by the in-plane pair Coulomb field. The calculations are carried out in the framework of the random phase approximation (RPA).

We have found that the static structure factor $S(q)$ as a function of the wave vector $q$ acquires a sharp peak around a certain value of $q = q_c$ when the PSOI parameter is large enough $\tilde{\alpha}/r_s \gtrsim 1/4 $, which indicates that the PSOI component of the e-e interaction is comparable in magnitude to the Fermi energy. The peak clearly shows the appearance of strong electron correlations on the $q_c$ scale, which are specific for the PSOI\@. They arise owing to the competition between the Coulomb repulsion of electrons and their attraction caused by the PSOI, which determines this characteristic spatial scale. Interestingly, the PSOI correlated state appears at rather high density of electrons and, correspondingly, at small $r_s$, when the usual Coulomb interaction is small.

When $\tilde{\alpha}/r_s$ exceeds a critical value, a new branch of solution with very unusual properties arises due to PSOI in the spectrum of collective excitations of the system in addition to common long-wave plasmons. The solution exists only in a finite band of wave vectors around $q_c$, the band width growing with $\tilde{\alpha}/r_s$. The solution frequency is purely imaginary, which means that the electron density fluctuations are growing with time. It is interesting that the density fluctuations are not polarized in spin. Thus the spatially uniform paramagnetic state of the electron system becomes unstable with respect to the charge density fluctuations on the $q_c$ scale. For realistic values of the SOI parameter $\tilde{\alpha} \ll 1$ in Rashba materials, the critical value $q_c \propto \tilde{\alpha}^{1/3} k_F$ lies in the long-wave part of the spectrum.

\section{Model and results}
In this section we consider the linear response of the 2D electron gas with PSOI to the external electric potential, the dynamic charge susceptibility, the static structure factor, and the spectrum of the collective modes. 

The 2D electron system is assumed to be symmetric with respect to the inversion of the normal to the plane. 
In this case the PSOI is produced by the in-plane pair Coulomb field in contrast to the gated one-dimensional quantum wire where only the normal component of the Coulomb field is important~\cite{PhysRevB.95.045138}. It is worth noting that the PSOI crucially depends on the geometry of the generating electric fields and momenta of interacting electrons. In the situation under consideration, both of these quantities are 2D vectors, the topology of which is determined self-consistently.  

The results are obtained using the equation of motion for the quantum Wigner function, which we derive and solve in the RPA, following Ref.~\cite{PhysRevB.95.045138}. The details of the calculation are presented in Appendix~\ref{appa}.

\subsection{Charge susceptibility}
The density $n^{(s)}_{q\omega}$ of the electrons with the $z$-component of the spin equal to $s=\pm 1$, in units of $\tfrac{\hbar}{2}$, satisfies the following system of linear equations
\begin{align}
\label{linearsystem}
    \chi_{0}^{-1} n^{(s)}_{q\omega} -  V_{q\omega}\sum_{\varsigma=\pm} n^{(\varsigma)}_{q\omega} = \varphi_{q\omega}\,,
\end{align}
with the external potential $\varphi_{q\omega}$, and the interaction potential
\begin{equation}
    V_{q \omega} = \mathcal{U}_q + 8 \frac{\alpha^2}{e^2} \mathcal{U}^2_q \chi_{j_{T}j_{T}}\,.
\end{equation}

The first term of the interaction potential is due to the Coulomb e-e repulsion. For the 2D electron gas formed in a uniform system with a bulk dielectric constant $\epsilon$ the e-e repulsion is governed by the pure Coulomb potential $\mathcal{U}_{q} = 2 \pi e^2 /\epsilon q$. The second term of the interaction potential is exactly due to the PSOI\@. The dynamic Lindhard susceptibility $\chi_{0}(q,\omega)$ and the transverse current-current susceptibility $ \chi_{j_{T}j_{T}}(q,\omega)$ are given by Eq.~\eqref{chi0} and Eq.~\eqref{chij}.

Since $V_{q \omega}$ is spin-independent, the solutions of Eq.~\eqref{linearsystem} correspond to the equal response of up- and down-spin densities, $n^{(+)}_{q\omega} = n^{(-)}_{q\omega}$. The dynamic charge susceptibility is
\begin{equation}
    \label{chin}
        \chi_{n}(q,\omega) =\frac{2 \chi_0(q,\omega)}{1 -2 \mathcal{U}_q [1 - \mathcal{G}_{\mathrm{PSOI}}(q,\omega)] \chi_0(q,\omega)}\,.
\end{equation}
The effect of the PSOI is described by the dynamic local field correction
\begin{equation}
        \mathcal{G}_{\mathrm{PSOI}}(q,\omega) = - \frac{8 \alpha^2}{e^2} \mathcal{U}_q\, \chi_{j_{T}j_{T}}(q,\omega)\,.
\end{equation}

\subsection{Static structure factor}
\begin{figure}[htp]
	\includegraphics[width=0.9\linewidth]{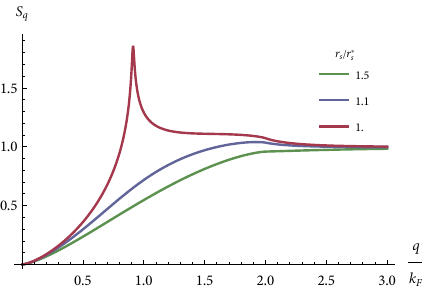}
	\caption{The static structure factor $S(q)$ as a function of $q$ for three values of the $r_s$. The PSOI magnitude is $\tilde{\alpha}=0.1$, which corresponds to $r_s^*=0.3$.\label{sfig}}
\end{figure}

Consider the static structure factor $S(q)$, which is related to the charge susceptibility of Eq.~\eqref{chin} via
\begin{equation}
    S(q) = -\frac{\hbar}{\pi n} \int_0^{\infty} d\omega\, \im \chi_n(q,\omega)\,,
\end{equation}
$n$ being the mean electron density. It is of interest to study the structure factor as a function of $q$ for different values of the e-e interaction parameters. Since there are two such parameters, it is convenient to fix the value of the PSOI constant $\tilde{\alpha}$ and change the parameter $r_s$ in such a way that both interaction parameters, $r_s$ and $\tilde{\alpha}/r_s$, are varied. The result is plotted in Fig.~\ref{sfig}. 

First of all, we found that the structure factor has a strong singularity at a certain value of the parameter $r_s=r_s^*$,
\begin{equation}
    r_s^* = \frac{2^{\frac{13}{6}} \tilde{\alpha}}{\sqrt{2^{\frac13} +2 {\tilde{\alpha}}^{\frac23}3^{\frac23}}}\,.
\end{equation}
As $r_s$ lowers down to this critical value, the spectral weight is shifted towards the long-wave part of the spectrum, and eventually a sharp peak is formed in the structure factor at the critical value $q_c$ of the wave-vector, given by
\begin{equation}
    \frac{q_c}{k_F} = \frac{2 {\tilde{\alpha}}^{\frac13} 3^{\frac13}}{\sqrt{2^{\frac13} +2 {\tilde{\alpha}}^{\frac23}3^{\frac23}}}\,,
\end{equation} 
which indicates the appearance of strong electron correlations due to PSOI\@.

\begin{figure}[htp]
	\includegraphics[width=0.9\linewidth]{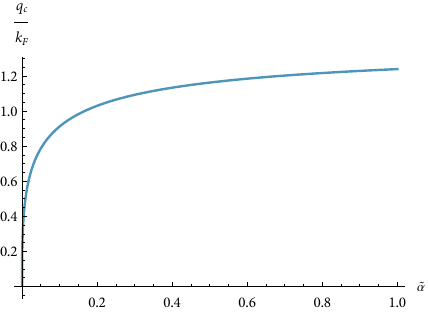}
	\caption{The critical value $q_c$ as a function of $\tilde{\alpha}$.\label{qcfig}}
\end{figure}

\begin{figure}[htp]
	\includegraphics[width=0.9\linewidth]{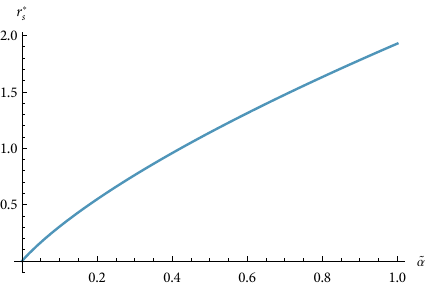}
	\caption{The critical value $r_s^*$ as a function of $\tilde{\alpha}$.\label{rsfig}}
\end{figure}

The characteristic spatial scale $q_c$ arises as a result of the competition between the Coulomb repulsion of electrons and their attraction caused by the PSOI\@. Its dependence on $\tilde{\alpha}$ is displayed in Fig.~\ref{qcfig}. When PSOI is extremely strong, $\tilde{\alpha} \gg 1$, the critical value tends to $q_c = \sqrt{2} k_F$. For small SOI parameter $\tilde{\alpha} \ll 1$, typically found in common Rashba materials, the critical value $q_c \propto \tilde{\alpha}^{1/3} k_F$ lies in the long-wave part of the spectrum. The dependence of the critical value of $r_s^*$ on the PSOI strength is plotted in Fig.~\ref{rsfig}.

\subsection{Collective modes and the instability of the uniform ground state}
The collective modes are given by zeroes of the denominator of Eq.~\eqref{chin}. This equation has two different types of solutions.

\begin{figure}[htp]
	\includegraphics[width=0.9\linewidth]{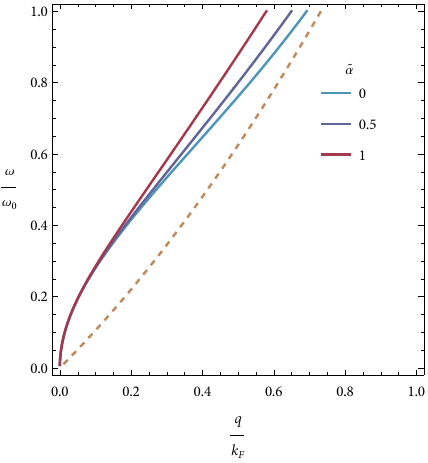}
	\caption{The plasmon frequency $\omega$ as a function of wave vector for three values of the PSOI constant, with $r_s=1$. The frequency is normalized at $\omega_0 = v_F k_F$. The dashed line shows the boundary of the particle-hole continuum.\label{figpl}}
\end{figure}

In the long-wave region the collective modes are common plasmons, the spectrum of which gets a correction from the PSOI\@. At $q \ll k_F$,
\begin{equation}
    \omega_{pl} = \omega_{2D} \sqrt{1+ \frac{\alpha^2}{e^2} q^2 k_F^2}\,,
\end{equation}
with
\begin{equation}
    \omega_{2D} = \sqrt{\frac{2 \pi e^2 n}{\epsilon m} q}
\end{equation}
being the standard plasmon dispersion in a 2D electron gas~\cite{PhysRevLett.18.546}. The correction, albeit small, can be accessible to experiment, since the high-accuracy measurements of the 2D plasmon spectra become available such as detecting the microwave absorption from the recombination photoluminescence spectrum~\cite{PhysRevB.102.081301,PhysRevB.105.L041403}. The plasmon spectrum is shown in Fig.~\ref{figpl}.

\begin{figure}[htp]
	\includegraphics[width=0.9\linewidth]{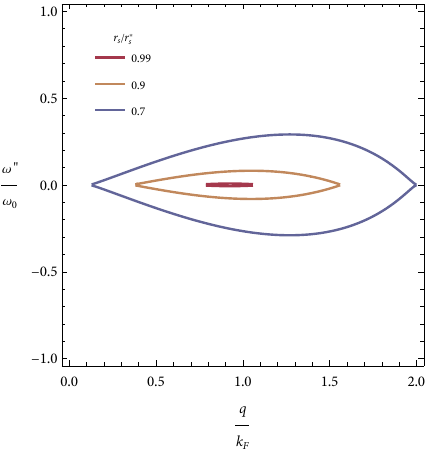}
	\caption{The imaginary part of the frequency of a new solution of the dispersion equation due to PSOI as a function of wave vector. The dispersion line is shown for three values of $r_s$ to trace how the instability develops in the system with increasing the PSOI interaction parameter of $\tilde{\alpha}/r_s$. The PSOI magnitude is fixed to be $\tilde{\alpha}=0.1$, which corresponds to $r_s^*=0.3$.\label{spect}}
\end{figure}

Most importantly, a new solution of the dispersion equation arises due to PSOI in addition to plasmons as soon as $r_s \le r_s^*$. At $r_s=r_s^*$ the solution appears to exist at a single critical value $q_c$ of the wave-vector, whereas at $r_s<r_s^*$ the region where the solution exists expands to a finite band of wave vectors $[q_1,q_2]$, the band width growing with lowering $r_s$. The spectrum of this branch of solution is illustrated by Fig.~\ref{spect}.

The frequency of the solution is purely imaginary. For every $q$ within the allowed band of $q \in (q_1,q_2)$ there are two branches, both with $\omega''>0$ and $\omega''<0$, forming together the petal-like shape. The frequencies of the two branches give, respectively, the increment and decrement of the time-dependent fluctuations in the system. The solution is characterized by equal response of up- and down-spin densities, $n^{(+)}_{q\omega} = n^{(-)}_{q\omega}$, which corresponds to the excitation in the charge sector. Hence the electron density fluctuations growing with time as $\propto \exp(\omega'' t)$ are not polarized in spin. This indicates that an instability of the spatially uniform paramagnetic ground state of the 2D electron gas with PSOI develops with respect to the charge density fluctuations on the $q_c$ scale. However, at $r_s>r_s^*$ the system is stable.

The effects of PSOI are enhanced in 2D atomically thin layers, either freely suspended~\cite{doi:10.1063/1.5019906,ROSSLER2010861} or immersed in a weak dielectric. The e-e repulsion is governed there by the Rytova-Keldysh potential
\begin{equation}
    \mathcal{U}_{q} = \frac{2 \pi e^2}{q (1 + q l)}\,,
\end{equation}
where $l$ is a characteristic length that can be roughly estimated as $\epsilon_{\parallel} d/2$, with $d$ being the layer thickness, $\epsilon_{\parallel}$ the in-plane dielectric constant of the layer material~\cite{RevModPhys.90.021001}. 

\begin{figure}[htp]
	\includegraphics[width=0.9\linewidth]{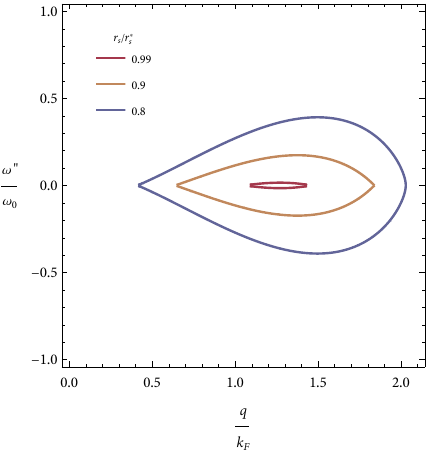}
	\caption{The imaginary part of the frequency of a new solution of the dispersion equation due to PSOI as a function of wave vector for three values of $r_s$. Here a 2D layer in vacuum is considered, its thickness being $d=0.02a_B$, the in-plane dielectric constant $\epsilon_{\parallel} = 15$. The PSOI magnitude is $\tilde{\alpha}=0.1$, which corresponds to $r_s^* = 1.054$. \label{spect_rk}}
\end{figure}

The spectrum of the new solution of the dispersion equation for this case is displayed in Fig.~\ref{spect_rk}. Because of the decreased dielectric screening the critical value $r_s^*$ increases as compared to the purely Coulomb case considered above.

\section{Concluding remarks}

We studied electron correlations and collective modes of the 2D electron system with strong PSOI produced by the in-plane Coulomb electric fields of interacting electrons, and come to the main conclusion that the PSOI leads to the formation of a specific strongly correlated state which becomes unstable with increasing the PSOI strength above a critical value.

The peculiarity of this system is that the e-e interaction strength is described by two interaction parameters, $r_s$ and $\tilde{\alpha}/r_s$, related to the Coulomb interaction and the PSOI\@. They are characterized by the opposite dependence on the electron density. The attractive PSOI prevails over the Coulomb repulsion when the density is high enough. 

This gives rise to strong electron correlations on a certain spatial scale that manifest themselves as a sharp peak in the static structure factor at $q=q_c$. Moreover, as soon as $\tilde{\alpha}/r_s$ exceeds a critical value, the dispersion equation defining the spectrum of the collective modes in the system features a new solution  with purely imaginary frequency existing in a band of wave vectors around $q_c$. In other words, a spatially uniform paramagnetic ground state becomes unstable with respect to the charge density fluctuations on the $q_c$ scale.

This altogether indicates a tendency for an electron state to form a striped structure of some sort on this spatial scale. Within the linear analysis undertaken in the present paper, it is impossible to predict a specific type of the electron state that would correspond to the true energy minimum. Instead of the perturbative RPA, the self-consistent approach, like the Hartree-Fock approximation~\cite{PhysRevB.85.035116,PhysRevB.54.1853}, is necessary to attack this problem.

The RPA, like any other perturbative approach, works well only for small values of the interaction parameter, which for the PSOI is $\tilde{\alpha}/r_s$. As can be seen from Fig.~\ref{rsfig}, at the instability threshold the PSOI interaction parameter is close to $1/2$. This is a reasonably good small parameter, so the RPA is justified for small $r_s$, when the Coulomb interaction is also small.

The obtained results, of course, do not imply that a sufficiently dense 2D electron gas with arbitrarily small PSOI is always unstable towards the density fluctuations. For very large electron density the considerations based on $k\cdot p$ approximation and the one-band model of SOI lose their validity when the Fermi energy is comparable to the gap energy from other Bloch bands involved in the formation of the PSOI Hamiltonian. Therefore the limiting value of $r_s$ is determined by the band structure of the specific material.

On the other hand, large values of $\tilde{\alpha} \gtrsim 1$ are not attainable at least in classical semiconductors with $sp^3$ band hybridization, where the upper limit for the Rashba constant is of the order of $\tilde{\alpha} \approx \epsilon^{-2}$, with $\epsilon$ being the bulk dielectric constant~\footnote{We are grateful to Sergei Tarasenko for pointing this out (S.~Tarasenko, private communication)}.

Nonetheless, the recent rise of 2D systems with giant SOI~\cite{doi:10.1021/acs.jpclett.1c03662,varignon2018new} gives us some hope for the realization of the extremely strong Rashba SOI by means of other physical mechanisms. Thus, in oxide heterostructures and films~\cite{Pai_2018,Stemmer_2018} the strong indications were found for the electronic nematicity~\cite{Levy_2020}, and for the formation of BEPs~\cite{Levy_2015,doi:10.1021/acs.nanolett.8b01614,mikheev2021clean} well beyond the superconducting phase. While the exact mechanisms behind these effects still remain unclear, at least some of them are likely due to the interplay of the giant SOI and collective effects~\cite{PhysRevB.104.125103}. That being said, the quest for a particular system with a giant SOI where the effects predicted in the present paper could develop is still a challenge for the future.

\begin{acknowledgments}
This work was carried out in the framework of the state task and supported by the Russian Foundation for Basic Research, Project No. 20--02--00126.
\end{acknowledgments}

\appendix
\begin{widetext}

\section{}
\label{appa}
In this section we obtain the linear response functions of the 2D electron gas with PSOI within RPA\@. The microscopic derivation is based on the equations of motion for the quantum Wigner distribution functions (WDFs)~\cite{hasegawa1975electron,PhysRevB.95.045138}.

We start with the second quantized form of the Hamiltonian~\eqref{ham}
\begin{align}
    \label{hm1}
    H ={}& -\frac{\hbar^2}{2m} \sum_{s} \int d\bm{r}\, \psi^+_{s}(\bm{r})\grad_{\bm{r}}^2 \psi_{s}(\bm{r})\\
    \label{hm2}
        &{}+ \frac{1}{2} \sum_{s_1 s_2} \int d\bm{r}_1 d\bm{r}_2\, \psi^+_{s_1}(\bm{r}_1) \psi^+_{s_2}(\bm{r}_2) \mathcal{U}(\bm{r}_1-\bm{r}_2) \psi_{s_2}(\bm{r}_2) \psi_{s_1}(\bm{r}_1)\\
        \label{hm3}
        &{} + \sum_{s} \int d\bm{r}\, \psi^+_{s}(\bm{r}) \varphi(\bm{r},t) \psi_{s}(\bm{r})\\ 
        \label{hm4}
    &{}+ \frac{2 m \alpha}{e \hbar} \int d\bm{r}_1 d\bm{r}_2 \normord{\hat{\rho}(\bm{r}_1) {\left( \bm{\mathcal{E}}(\bm{r}_1 - \bm{r}_2) \times  \hat{\bm{j}}_{\sigma}(\bm{r}_2)\right)}_z}\,,
\end{align} 
where $\psi_{s}(\bm{r})$ is the electron field operator in the $s_z$ representation ($s = \pm 1$), $\bm{r} \equiv (x,y,0)$ stands for the in-plane position, the colons stand for the normal ordering, $\hat{\rho}(\bm{r})$ is the density operator, and $\hat{\bm{j}}_{\sigma}(\bm{r})$ is the paramagnetic part of the spin current,
\begin{align}
    \hat{\bm{j}}_{\sigma}(\bm{r}) = \frac{i e \hbar}{2m} \sum_{\varsigma} \varsigma \left[\hat{\psi}^{+}_{\varsigma}(\bm{r}) \nabla_{\!\bm{r}} \hat{\psi}_{\varsigma}(\bm{r}) - (\nabla_{\!\bm{r}} \hat{\psi}^{+}_{\varsigma}(\bm{r})) \hat{\psi}_{\varsigma}(\bm{r})\right]\,.
\end{align}
Eq.~\eqref{hm1} is the kinetic energy operator, Eq.~\eqref{hm2} is the Coulomb interaction, Eq.~\eqref{hm3} is the external perturbation, and Eq.~\eqref{hm4} is the PSOI\@.

Let us introduce the single-particle and two-particle Klimontovich operators~\cite{klimontovich} as
\begin{equation}
	\hat{f}^{(s)}(\bm{r},\bm{p},t) = \frac{1}{{(2 \pi)}^2} \int d\bm{R}\, e^{i \bm{p} \cdot \bm{R}}\psi_s^+\left(\bm{r} + \tfrac{\bm{R}}{2},t\right)\psi_s\left(\bm{r} - \tfrac{\bm{R}}{2},t\right)
\end{equation}
and
    \begin{equation}
        \hat{f}^{(s_1, s_2)}(\bm{r}_1,\bm{p}_1,\bm{r}_2,\bm{p}_2,t) = 
        \frac{1}{{(2 \pi)}^4} \int d\bm{R}_1 d\bm{R}_2\, e^{i (\bm{p}_1 \cdot\bm{R}_1 + \bm{p}_2 \cdot\bm{R}_2)} \psi_{s_1}^+\left(\bm{r}_1 + \tfrac{\bm{R}_1}{2},t\right)\psi_{s_2}^+\left(\bm{r}_2 + \tfrac{\bm{R}_2}{2},t\right)\psi_{s_2}\left(\bm{r}_2 - \tfrac{\bm{R}_2}{2},t\right)\psi_{s_1}\left(\bm{r}_1 - \tfrac{\bm{R}_1}{2},t\right)\,.
    \end{equation}
The expectation value of the Klimontovich operators w.r.t.\ the ground state of $H$ are exactly the WDFs, i.e.\ $\av*{\hat{f}^{(s)}(\bm{r},\bm{p},t)} \equiv {f}^{(s)}(\bm{r},\bm{p},t)$. By commuting $\hat{f}^{(s)}(\bm{r},\bm{p},t)$ with $H$ and taking the expectation value we obtain the first equation of motion in the Bogoliubov-Born-Green-Kirkwood-Yvon hierarchy~\cite{bonitz2016quantum}:
\begin{align}
\label{BBGKY}
        i \hbar \partial_t f^{(s)}(\bm{r},\bm{p},t) ={} &- \frac{i \hbar^2}{m}\bm{p} \cdot \grad_{\bm{r}} f^{(s)}(\bm{r},\bm{p},t) + \frac{1}{{(2\pi)}^2} \int d \bm{R} d\bm{p}_1\, e^{i(\bm{p} - \bm{p}_1)\cdot \bm{R}} \left\{\varphi\left(\bm{r} - \tfrac{\bm{R}}{2},t\right) - \varphi\left(\bm{r} + \tfrac{\bm{R}}{2},t\right)\right\} f^{(s)}(\bm{r},\bm{p}_1,t)\\
        &{}+ \frac{1}{{(2\pi)}^2} \sum_{\varsigma}\int d\bm{r}_1 d\bm{R}\, d\bm{p}_1 d\bm{p}_2\, e^{i(\bm{p} - \bm{p}_1)\cdot \bm{R}}  \left\{\mathcal{U} \left(\bm{r} - \bm{r}_1 - \tfrac{\bm{R}}{2}\right) - \mathcal{U} \left(\bm{r} - \bm{r}_1 + \tfrac{\bm{R}}{2}\right)\right\}f^{(s, \varsigma)}(\bm{r},\bm{p}_1,\bm{r}_1,\bm{p}_2,t)\notag\\
        &{}- \frac{i \alpha}{2 \pi^2} \sum_{\varsigma} \int d\bm{r}_1 d\bm{R}\, d\bm{p}_1 d\bm{p}_2\, e^{i(\bm{p} - \bm{p}_1)\cdot \bm{R}} \mathcal{E}_y\left(\bm{r}_1 - \bm{r} + \tfrac{\bm{R}}{2}\right)\left[\varsigma(\tfrac12 \partial_{x_1}+ip_{2x}) - s(\tfrac12 \partial_{x}+ip_{1x})\right] f^{(s, \varsigma)}(\bm{r},\bm{p}_1,\bm{r}_1,\bm{p}_2,t)\notag\\
        &{}+\frac{i \alpha}{2 \pi^2} \sum_{\varsigma} \int d\bm{r}_1 d\bm{R}\, d\bm{p}_1 d\bm{p}_2\, e^{i(\bm{p} - \bm{p}_1)\cdot \bm{R}}\mathcal{E}_x\left(\bm{r}_1 - \bm{r} + \tfrac{\bm{R}}{2}\right)\left[\varsigma(\tfrac12 \partial_{y_1}+ip_{2y}) - s(\tfrac12 \partial_{y}+ip_{1y})\right] f^{(s, \varsigma)}(\bm{r},\bm{p}_1,\bm{r}_1,\bm{p}_2,t)\notag\\
        &{}-\frac{i \alpha}{2 \pi^2} \sum_{\varsigma} \int d\bm{r}_1 d\bm{R}\, d\bm{p}_1 d\bm{p}_2\, e^{i(\bm{p} - \bm{p}_1)\cdot \bm{R}}\mathcal{E}_y\left(\bm{r}_1 - \bm{r} - \tfrac{\bm{R}}{2}\right)\left[\varsigma(\tfrac12 \partial_{x_1}-ip_{2x}) - s(\tfrac12 \partial_{x}-ip_{1x})\right] f^{(s, \varsigma)}(\bm{r},\bm{p}_1,\bm{r}_1,\bm{p}_2,t)\notag\\
        &{}+\frac{i \alpha}{2 \pi^2} \sum_{\varsigma} \int d\bm{r}_1 d\bm{R}\, d\bm{p}_1 d\bm{p}_2\, e^{i(\bm{p} - \bm{p}_1)\cdot \bm{R}}\mathcal{E}_x\left(\bm{r}_1 - \bm{r} - \tfrac{\bm{R}}{2}\right)\left[\varsigma(\tfrac12 \partial_{y_1}-ip_{2y}) - s(\tfrac12 \partial_{y}-ip_{1y})\right] f^{(s, \varsigma)}(\bm{r},\bm{p}_1,\bm{r}_1,\bm{p}_2,t)\,.\notag
\end{align} 
RPA consists in truncating the hierarchy by factorizing the two-particle WDF according to
\begin{equation}
	f^{(s_1, s_2)}(\bm{r}_1,\bm{p}_1,\bm{r}_2,\bm{p}_2,t) = f^{(s_1)}(\bm{r}_1,\bm{p}_1,t)f^{(s_2)}(\bm{r}_2,\bm{p}_2,t)\,.
\end{equation}
Let us define the deviation $f^{(s)}_1(\bm{r},\bm{p},t)$ of $f^{(s)}(\bm{r},\bm{p},t)$ from its equilibrium value $f_0^{(s)}(\bm{p})$ as a result of the external perturbation,
\begin{equation}
	f^{(s)}_1(\bm{r},\bm{p},t) = f^{(s)}(\bm{r},\bm{p},t) - f^{(s)}_0(\bm{p})\,.
\end{equation}

The equation of motion for $f^{(s)}_1(\bm{r},\bm{p},t)$, linearized with respect to the external potential, reads in Fourier representation as
\begin{equation}
\label{eom}
\begin{split}
    - \hbar\omega f^{(s)}_1(\bm{q},\bm{p},\omega) ={}& -\frac{\hbar^2 \bm{p}\cdot \bm{q}}{m}f^{(s)}_1(\bm{q},\bm{p},\omega) + \left[f^{(s)}_0(\bm{p}+\tfrac{\bm{q}}{2}) - f^{(s)}_0(\bm{p}-\tfrac{\bm{q}}{2})\right] \times \\
    &\times \left\{ \varphi_{q \omega} + \mathcal{U}_{q} \sum_{\varsigma} n^{(\varsigma)}_{\bm{q} \omega} + 2\alpha s {(\bm{p} \times \bm{\mathcal{E}}_{\bm{q}})}_{z}  \sum_{\varsigma} n^{(\varsigma)}_{\bm{q}\omega} + 2 \alpha\frac{m}{e \hbar} \sum_{\varsigma}\varsigma {(\bm{j}^{(\varsigma)}_{\bm{q}\omega}\times \bm{\mathcal{E}}_{\bm{q}})}_{z} \right\} \,.
\end{split}
\end{equation}
\end{widetext}

Here $n^{(s)}_{q \omega}$ and $\bm{j}^{(s)}_{\bm{q}\omega}$ are the $s$-spin components of the electron density and paramagnetic current response, related to the WDF, respectively, by 
\begin{equation}
    n^{(s)}_{q \omega}  = \int d\bm{p}\, f^{(s)}_1(\bm{q},\bm{p},\omega)
\end{equation}
and
\begin{equation}
    \bm{j}^{(s)}_{\bm{q}\omega}  = -\frac{e \hbar}{m}\int d\bm{p}\, \bm{p} f^{(s)}_1(\bm{q},\bm{p},\omega)\,.
\end{equation}
The Fourier component of the in-plane electric field is given by $\bm{\mathcal{E}}_{\bm{q}} = \frac{\bm{q}}{i e}\mathcal{U}_{q}$. Two specific forms of the e-e repulsion potential $\mathcal{U}_{q}$ are considered in the main text.

The first two moments of the Eq.~\eqref{eom} are
\begin{align}
    n^{(s)}_{q \omega} ={}& \chi_{0}\left(\varphi_{q \omega} + \mathcal{U}_{q} \sum_{\varsigma} n^{(\varsigma)}_{q \omega} + 2 \alpha\frac{m}{e \hbar} \sum_{\varsigma}\varsigma {(\bm{j}^{(\varsigma)}_{\bm{q}\omega}\times \bm{\mathcal{E}}_{\bm{q}})}_{z}\right)\notag \\  
    \label{mom0}
    &{}+ \chi_{1}  \frac{2\alpha s}{i e} \mathcal{U}_{q}  \sum_{\varsigma} n^{(\varsigma)}_{q \omega}\,,
\end{align}
and
\begin{align}
    -\frac{m}{e \hbar}\bm{j}^{(s)}_{\bm{q}\omega} ={}& \bm{\chi}_{2}\left(\varphi_{q \omega} + \mathcal{U}_{q} \sum_{\varsigma} n^{(\varsigma)}_{q \omega} + 2 \alpha\frac{m}{e \hbar} \sum_{\varsigma}\varsigma {(\bm{j}^{(\varsigma)}_{\bm{q}\omega}\times \bm{\mathcal{E}}_{\bm{q}})}_{z}\right)\notag\\
    \label{mom1}
    &{} + \bm{\chi}_{3}  \frac{2\alpha s}{i e} \mathcal{U}_{q}  \sum_{\varsigma} n^{(\varsigma)}_{q \omega}\,.
\end{align}
Four susceptibilities entering the Eqs.~(\ref{mom0}--\ref{mom1}) are as follows,
\begin{align}
    \chi_{0} &= \int d\bm{p}\, \frac{f^{(s)}_0(\bm{p}+\frac{\bm{q}}{2}) - f^{(s)}_0(\bm{p}-\frac{\bm{q}}{2})}{\frac{\hbar^2}{m}\bm{p}\cdot\bm{q} - \hbar (\omega +i0)}\,,\\
    \chi_{1} &= \int d\bm{p}\, \frac{f^{(s)}_0(\bm{p}+\frac{\bm{q}}{2}) - f^{(s)}_0(\bm{p}-\frac{\bm{q}}{2})}{\frac{\hbar^2}{m}\bm{p}\cdot\bm{q} - \hbar (\omega +i0)}{(\bm{p}\times \bm{q})}_z\,,\\
    \bm{\chi}_{2} &= \int d\bm{p}\, \frac{f^{(s)}_0(\bm{p}+\frac{\bm{q}}{2}) - f^{(s)}_0(\bm{p}-\frac{\bm{q}}{2})}{\frac{\hbar^2}{m}\bm{p}\cdot\bm{q} - \hbar (\omega +i0)} \bm{p}\,,\\
    \bm{\chi}_{3} &= \int d\bm{p}\, \frac{f^{(s)}_0(\bm{p}+\frac{\bm{q}}{2}) - f^{(s)}_0(\bm{p}-\frac{\bm{q}}{2})}{\frac{\hbar^2}{m}\bm{p}\cdot\bm{q} - \hbar (\omega +i0)}\bm{p}{(\bm{p}\times \bm{q})}_z\,. 
\end{align}
Note that $\bm{\chi}_{2}$ and $\bm{\chi}_{3}$ are vector quantities.

In a non-self-consistent RPA, $f^{(s)}_0(\bm{p})$ is the momentum distribution of the free 2D paramagnetic electron gas, 
\begin{equation}
    f^{(s)}_0(\bm{p}) = \frac{1}{{(2 \pi)}^2} \theta(k_F-|\bm{p}|)\,,
\end{equation}
$k_F$ being the Fermi wave vector. In this case
\begin{widetext}
\begin{align}
    \label{chi0}
    \chi_{0} ={}& \frac{N_{\sigma}}{(q/k_F)} \left( \nu_{-} - \sign (\re \nu_{-}) \sqrt{\nu_{-}^2-1} -\nu_{+} + \sign (\re \nu_{+}) \sqrt{\nu_{+}^2-1}\right)\,,\\
    \chi_{1} ={}& 0\,,\\
    \bm{\chi}_{2} ={}& \frac{\bm{q} N_{\sigma}}{2(q/k_F)} \left( \nu_{-} - \sign (\re \nu_{-}) \sqrt{\nu_{-}^2-1} +\nu_{+} - \sign (\re \nu_{+}) \sqrt{\nu_{+}^2-1}\right)\\
    &{}+\frac{\bm{q} N_{\sigma}}{{(q/k_F)}^2} \left( \nu_{-}^2 - \nu_{-}\sign (\re \nu_{-}) \sqrt{\nu_{-}^2-1} -\nu_{+}^2 + \nu_{+}\sign (\re \nu_{+}) \sqrt{\nu_{+}^2-1}\right)\,,\\
    \bm{\chi}_{3} ={}& N_{\sigma} k_F^2 \frac{\bm{q}_{\perp}}{(q/k_F)} \left( \frac{\nu_{-}}{2} - \frac13 \sign (\re \nu_{-}) \left[{(\nu_{-}^2)}^{\frac32} - {(\nu_{-}^2-1)}^{\frac32}\right] - \frac{\nu_{+}}{2} + \frac13 \sign (\re \nu_{+}) \left[{(\nu_{+}^2)}^{\frac32} - {(\nu_{+}^2-1)}^{\frac32}\right] \right)\,.
\end{align}
Here $\nu_{\pm} = \frac{\omega +i0}{q v_F} \pm \frac{q}{2k_F}$, $N_{\sigma} = \frac{m}{2 \pi \hbar^2}$ is the 2D density of states of spin $\sigma$ per unit area, $\bm{q}_{\perp} \equiv (q_y,-q_x) $, the branch cut lies along the negative real axis. Eq.~\eqref{chi0} is due to Stern~\cite{PhysRevLett.18.546}.

Expressing $n^{(s)}_{q \omega}$ from a coupled system of equations Eqs.~(\ref{mom0}--\ref{mom1}) gives Eq.~\eqref{linearsystem} of the main text with
\begin{align}
    \label{chij}
    \chi_{j_{T}j_{T}} \equiv {} {(\bm{\chi}_{3} \times \bm{q})}_z 
    = N_{\sigma} k_F^4 \frac{q}{k_F} \Bigg( \frac{\nu_{-}}{2} - \frac13 \sign (\re \nu_{-}) \left[{(\nu_{-}^2)}^{\frac32} - {(\nu_{-}^2-1)}^{\frac32}\right] - \frac{\nu_{+}}{2} + \frac13 \sign (\re \nu_{+}) \left[{(\nu_{+}^2)}^{\frac32} - {(\nu_{+}^2-1)}^{\frac32}\right] \Bigg)\,.
\end{align}
\end{widetext}

\bibliography{paper}
\end{document}